# The penetration of plasma clouds across magnetic boundaries: the role of high frequency oscillations for magnetic diffusion

T. Hurtig, N. Brenning, and M. A. Raadu

Alfvén Laboratory, Royal Institute of Technology, se-100 44 Stockholm, Sweden

**Abstract**

Experiments are reported where a collisionfree plasma cloud penetrates a magnetic barrier by self-polarization. We here focus on the resulting anomalous magnetic field diffusion into the plasma cloud, two orders of magnitude faster than classical, which is one important aspect of the plasma cloud penetration mechanism. Without such fast magnetic diffusion, clouds with kinetic $\beta_k$ below unity would not be able to penetrate magnetic barriers at all. Tailor-made diagnostics has been used for measurements in the parameter range with the kinetic $\beta_k \approx 0.5$ to 10, and with normalized width $w/r_{gi}$ of the order of unity. Experimental data on hf fluctuations in density and in electric field has been combined to yield the effective anomalous transverse resistivity $\eta_{EFF}$. It is concluded that they are both dominated by highly nonlinear oscillations in the lower hybrid range, driven by a strong diamagnetic current loop that is set up in the plasma in the penetration process. The anomalous magnetic diffusion rate, calculated from the resistivity $\eta_{EFF}$, is consistent with single-shot multi-probe array measurements of the diamagnetic cavity and the associated quasi-dc electric structure. An interpretation of the instability measurements in terms of the resistive term in the generalized (low frequency) Ohm's law is given.

## 1. Introduction.

Plasmoid penetration across magnetic barriers has been the subject of a series of four recent papers in Physics of Plasmas from our group [1-4]. Several interlocking processes are important for the penetration mechanism. Here, we will focus on only one of them, anomalous fast magnetic field diffusion. For the complete picture, the reader is referred to papers [1–4]. Fig. 1 shows the magnetic diffusion mechanism schematically in a case where it is comparatively slow, so that different phases can be distinguished. A plasmoid (a fast plasma cloud) is here assumed to have penetrated into a region with a transverse magnetic field. The magnetic field diffusion into the plasmoid goes though three phases. First, a diamagnetic phase where the transverse magnetic component is excluded from the plasmoid by a diamagnetic current loop around the edge. Second, a diffusion phase where the magnetic field diffuses into the cloud and, third, a propagation phase when the magnetic field has fully diffused into the plasmoid, which is able to continue its motion because it has self-polarized to the **E** = - **vxB** field strength.

We have made experiments in the plasma gun device shown in the top panel of Fig. 2. With a plasma speed of $3 \times 10^5$ m/s and a transition region of 0,3m length, the plasma crosses the magnetic transition in about one microsecond. For typical plasma stream parameters (density $3 \times 10^{18}$ m$^{-3}$, stream width 0,1m, and electron temperature 6-8eV), the classical magnetic diffusion time based on the Spitzer resistivity is of the order of 100 microseconds [3]. The plasma would move 30 meters during this time. One would therefore, if the magnetic diffusion were classical, expect the transverse component of the magnetic field to be almost completely excluded from the interior of the plasmoid in the whole experimental device of Fig. 2, which extends less than one meter downstream from the magnetic transition.

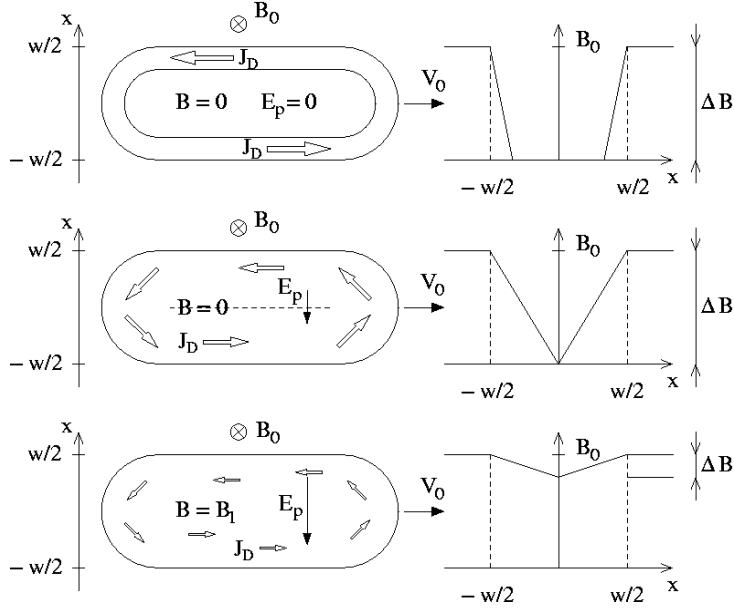

Fig. 1. The penetration of magnetic fields into a moving plasma cloud (or plasmoid) in three phases. (a) Diamagnetic exclusion of the field. (b) Diffusion of the field. (c) Continued plasmoid propagation by the **ExB** drift.

This is not what we observe. The middle panel of Fig. 3 shows the magnetic field component $B_y$ as function of time, measured by a multi-probe array spaced along the $x$ axis at $y = z = 0$, immediately after the magnetic transition [3]. The transverse ($B_y$) component has already here penetrated the plasmoid by about 70% (notice that the diamagnetic reduction of the magnetic field inside the plasmoid shows up as a peak in Fig. 3 due to the negative sign of $B_y$). The potential in the top panel of Fig. 3 agrees with the self-polarization field of Fig. 1c. The plasmoid in Fig. 3 is thus in a stage somewhere between the diffusion and propagation phases of Fig. 1. We conclude that the magnetic diffusion time must be of the same order as the transition time, about 1 μs, and thus about a factor 100 faster than classical. ( A comment might be necessary here. We claim that magnetic field diffusion of some kind is necessary for a plasmoid to enter a region with a transverse magnetic field, in such a fashion that the plasmoid keeps is general shape, velocity, and orientation, and the transverse field penetrates the plasmoid. We motivate this claim in the rest frame of the plasmoid. In this frame, the plasma cloud is stationary and is subject to a time-changing transverse magnetic field. For the magnetic field to penetrate the plasma, some magnetic diffusion process is clearly necessary).

If classical magnetic diffusion is too slow, some anomalous process is needed. There are two fundamentally different types of mechanisms which can give anomalous fast magnetic penetration into a plasmoid: violation of the frozen-in condition due to parallel (magnetic-field-aligned) electric fields, or enhanced magnetic diffusion on the micro-scale, by waves or turbulence. Both types are reported in the literature but the question when one, or the other type, dominates remains to be resolved. The idea that parallel electric fields can arise in the vicinity of cross-**B**-moving plasmoids was first proposed in connection with ionospheric injection experiments [5,6,7]. Parallel electric fields on the edges of a moving plasma cloud were proposed to partly screen the internal self-polarization field **E** = -**v**x**B** from the ambient medium. This weakens the frozen-in condition and allows a plasmoid injected in the ionosphere to move across **B** without dragging along all ambient plasma in the magnetic flux tube. This is the mechanism invoked by Delamaire *et al* [5] to explain anomalous long skidding distances observed in the ionospheric CRRES releases of Barium clouds. Two different mechanisms, by which moving clouds should be able to maintain such parallel fields, have been proposed by Brenning [6], and by Brenning and Fälthammar [7]. Although these papers [5,6,7] do not directly treat magnetic diffusion into plasmoids, they give some arguments for the establishment of parallel electric fields around moving plasmoids. The direct role of parallel electric fields in plasmoid penetration across a magnetic barrier (and thus, implicitly, magnetic field diffusion) was to our knowledge first demonstrated in [1] where we presented computer simulations in a parameter regime where waves and turbulence do not have time to grow. In a recent work, Echim [8] has proposed that parallel electric fields would allow clumps in the solar wind plasma to enter the Earth's magnetosphere.

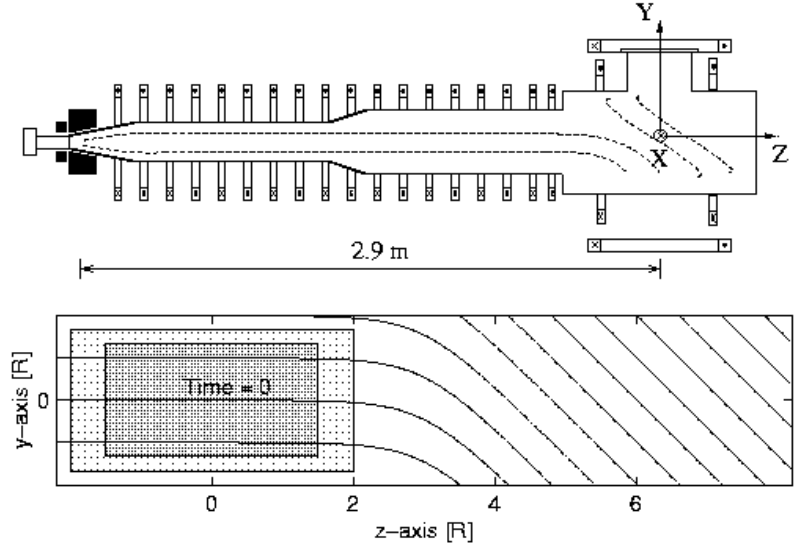

Fig. 2. Top panel: a schematic picture of the experiment. Bottom panel: the simulation region of [2], where the simulation region is shown in light grey and the injected plasma in dark grey.

The role of waves, or turbulence of some kind, to enable anomalous magnetic diffusion is more established in the literature and will therefore only be briefly mentioned here. Different mechanisms, for example whistler waves, the modified two-stream instability (MTSI), and the lower hybrid drift instability (LHDI) have been invoked in this context. A recent paper, which advocates whistler waves, is [9]. References can be found in [3,9,10].

In the present work we present data on the magnetic diffusion mechanism in one experiment, shown in Fig. 2. The approach is that of the experimentalist rather than the theoretician. Without assuming anything about the physical nature of the waves in the experiment, or about the driving mechanism behind, we will directly from measured wave data demonstrate their role in the magnetic diffusion process. In another publication [3] we have shown that the modified two-stream instability, MTSI, driven by a diamagnetic current loop, agrees quite well with the wave data and probably is the operating mechanism in this particular case.

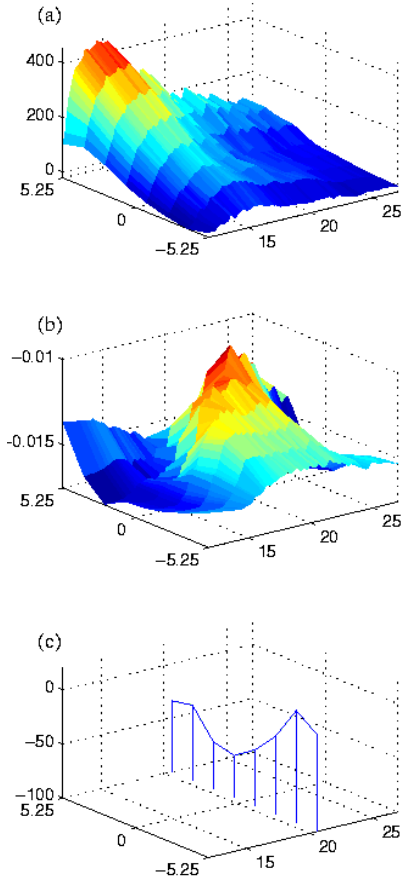

Fig. 3. Measurements with multi-probe arrays across the plasma beam, along the *x*-axis, at $z = y = 0$. Top plate (a): The potential in the stationary laboratory frame as a function of time. Middle plate (b): The transverse magnetic field component $B_y$ as function of time. Bottom plate (c): The potential in the moving rest frame of the plasma.

## 2. The experiment and diagnostics

Our plasma gun is shown in Fig. 2 and described in [3]. Typical parameters are: $B$ = 0,015T, $n_e$ = $10^{18}$-$10^{19}$m$^{-3}$, $T_e$ = 5-10eV, $W_i$ = 200 – 2000eV, stream velocity 2 - 6x$10^5$m/s, and stream width $w$ = 0,1m. The ion (hydrogen) gyro radius at typical stream velocities is 0,1 – 0,3 m. The diagnostics is of two types , high and low frequency. The low frequency (< 2MHz) diagnostics consists of Rogowski coils for currents, and of multi probe arrays (for electric and magnetic fields), which are used to obtain single-shot space resolved data in time series which cover the whole entry of the plasmoid across the magnetic barrier. This is the type of data in Fig. 3, and it shows the quasi-dc features of the process: the establishment of a self-polarization **E** = -**v**x**B** field (Fig. 3a), the magnetic diffusion into the plasmoid (Fig. 3b), and the internal electric structure in the plasma rest frame (Fig. 3c). The Rogowski coils are used in parallel to the magnetic measurements to map the large-scale diamagnetic current system, schematically shown in Fig. 4a. This diamagnetic current is quite strong, of the order of $10^5$ Am$^{-2}$, and is established in the magnetic transition region. It drives weaker currents upstream of the transition, and is partly frozen into the plasmoid downstream of the transition, as show in Fig. 4, and discussed in detail in [2].

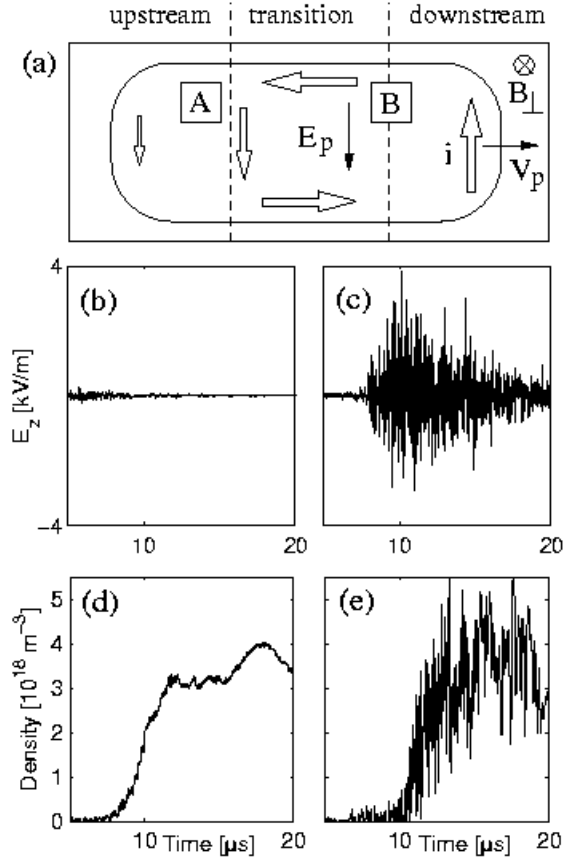

Fig. 4. (a) A schematic picture of the currents and the polarization field $E_p$ = - $\mathbf{v} \times \mathbf{B}$ in the plasma gun. (b, d) The flow-aligned high frequency component $E_{wz}$ of the electric field, and the density, at position A. (c, e) The flow-aligned high frequency component $E_{wz}$ of the electric field, and the density, at position B.

The high frequency diagnostics consist of electric and magnetic probes. They are used for wave features above about 2Mhz that are filtered away in the low frequency diagnostics. In addition there are plasma density measurements that cover the whole range from dc to several 10's of MHz. The high frequency signals, both in density and electric field, have similar spectra with a half width of about 6-7MHz and maximum amplitude at 7-8MHz, as shown in Fig. 5. This maximum is slightly below the lower hybrid frequency $\omega_{lh} = (\omega_{gi}\omega_{ge})^{1/2} \approx$ 10MHz.

## 3. Measurements on the hf features, and magnetic diffusion

It would have been desirable (and is planned) to make careful studies of the hf waves in the whole plasma stream, and as function of time. This is however a large programme. As a first step we have studied the waves carefully around only one position, and around one time window of 2 µs. The choice of position is $(x, y, z) = (0,03, 0, 0)$, denoted by B in Fig. 4. We will call this the standard position B. This is right in the middle of the diamagnetic current sheath, which flows along the plasma stream on the high potential side (*i e*, the positive *x* side). At this *x* coordinate there is the steepest slope d$B$/d$x$ of the diamagnetic magnetic cavity of Fig. 3b, and consequently this is where we best should see signs of the process that gives magnetic diffusion. We want to have a look at the magnetic diffusion in the *x* direction, from the sides into the plasmoid, and therefore want to avoid the front and back of the plasmoid. The time window we chose is around 13 µs after the shot of the plasma gun, by the following compromise. On one hand, we do not want to take too early a time, which would put us in the front of the plasma stream. On the other hand, if we take too late a time, plasma will have reached the bottom plate (at $z = 0{,}5$m), and begun to pile up against the plasma stream. Fig 4e shows the plasma density as function of time at $z = 0$. Around 13 µs the front has passed the position of measurement ($z = 0$), and just marginally has had time to reach the bottom plate.

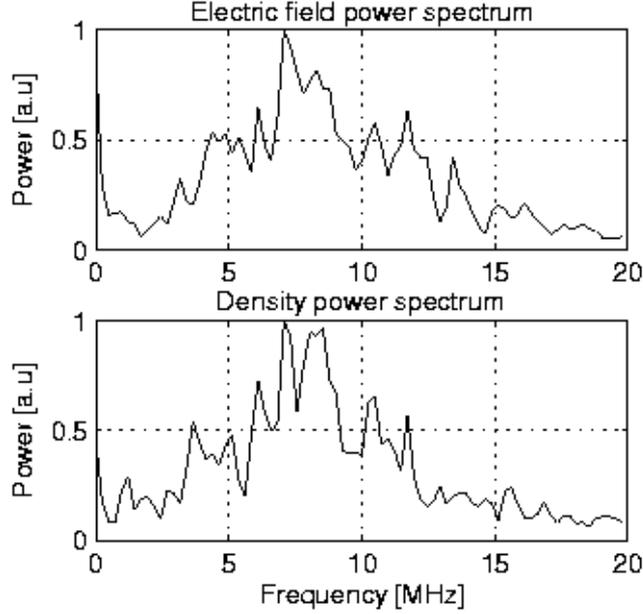

fig. 5. Frequency spectra for the flow aligned wave ($E_{wz}$) electric field (top) and the density oscillations (bottom). The data is averaged over ten shots. For reference, the lower hybrid frequency is 10 MHz.

The key high frequency measurements at the standard position B are the plasma-flow-aligned $E_z$ component, and the plasma density. Fig. 4b and 4c show that the hf amplitude in $E_z$ increases dramatically from position A to position B, reaching several kV/m when the plasma has entered the transition region. Also the plasma density, which is rather oscillation-free at A, before the transition (Fig. 4d), shows high amplitude oscillations after the transition, at position B (Fig. 4e).

Figure 6 shows details in a time window of 2 µs of one shot, taken at the standard position B. In the hf part of the electric field, the component $E_{wz}$ in the flow direction dominates totally, as can be seen by comparing the top three panels. For the discussion here, we disregard the other two components of the hf $\mathbf{E}_W$ field. Measurements with small arrays of hf probes around the standard position B yields data for these waves [3,4]: the typical wavelength is 30 mm, and thus is clearly resolved by the hf probes which have a spatial resolution of 5 mm. The wave vector **k** is oriented closely along the flow direction of the plasma stream. There is strong spatial coherence in both $E_{wz}$ and density, extending some cm in both the *x* and *y* directions. The typical amplitude in the density oscillations are usually about 50%, but sometimes reaches 100% as in the example shown in Fig. 6d. Thus the waves can (at the standard position B) be regarded as shown in Fig. 7 which is drawn approximately to scale: slabs of increased plasma density, with 30mm distance between the maxima, and oriented across the plasma flow direction. They are associated with hf $E_{wz}$ wave fields, which oscillate back and forth in the $z$ direction (Fig. 7c), and (from three-axis magnetic pickup coils, not shown here) with hf currents back and forth in the $x$ direction (Fig. 7d).

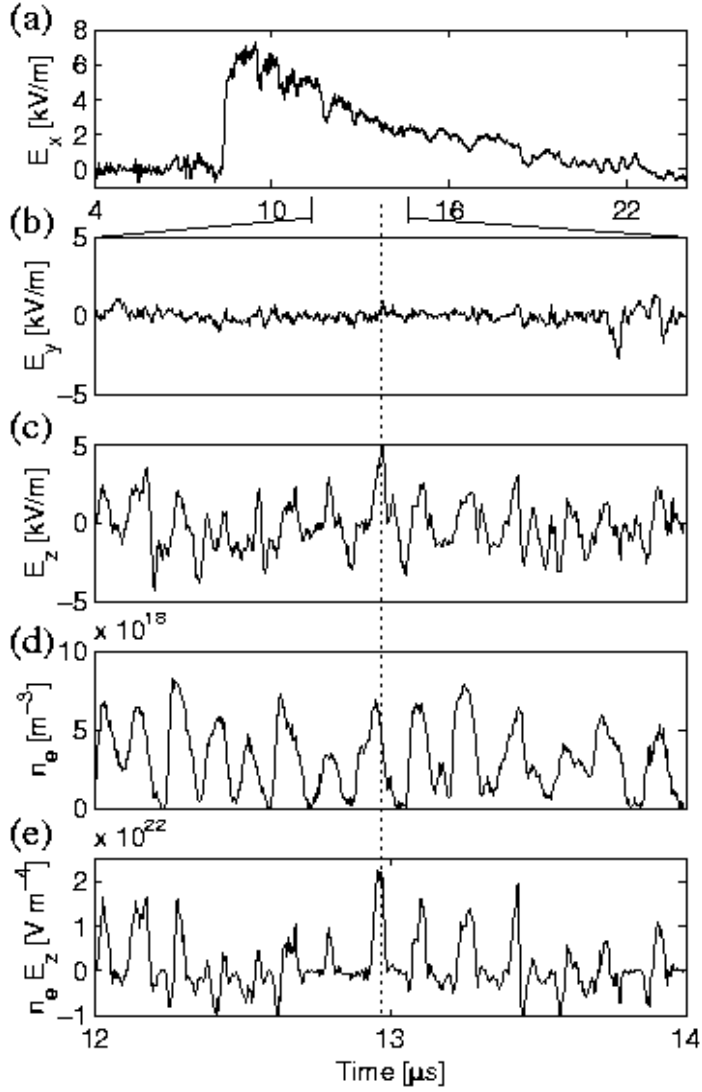

Fig. 6. (a) A 20 μs recording of the polarisation field $E_x$. (b) 2 μs of the hf fields $E_{wy}$ at position B of Fig. 4. (c) 2 μs of the hf field $E_{wz}$ at position B of Fig. 4, on the same scale. (d) A density recording at the same position. (d) The product of density and $E_z$.

The wave phase velocity $v_{wz}$ is measured by probe pairs at different $z$ coordinates, by varying the delay time to maximize the cross correlation. It is, at the standard position B, of the order of $10^5$ ms$^{-1}$ above the ion velocity $v_{iz}$ as deduced from time of flight from the plasma gun, but below the electron velocity $v_{ez}$ in the $z$ direction. The electron velocity $v_{ez}$ has here [3] been deduced in two independent ways which give consistent results, (1) from the measured diamagnetic current density $i_{Dz} = \langle n_e e \rangle (v_{iz} - v_{ez})$, and (2) from the electron drift in the macroscopic fields **E** and **B**, taken locally and at the proper time from measurements such as in Fig. 3a and 3b.

We will now have a look, in physical terms, at how the wave data in the third and fourth panel of Fig. 6 can be related to anomalous magnetic field diffusion, and then in the next section discuss the physics in the context of the generalized Ohm's law. We notice three features regarding the waves:

(1) The dominating hf electric field component $E_{wz}$ is in the same direction as (or to be precise, oscillating between parallel and antiparallel to) the macroscopic diamagnetic current, which flows in the negative $z$ direction at the standard position B in Fig. 4 where (and the time when) these waves are measured.

(2) The time series of the product $n_e E_{wz}$ is calculated and plotted in the bottom panel of Fig. 6. The $E_{wz}$ field and the density are clearly correlated in such a fashion that the maxima in density coincide with $E_{wz}$ fields in the positive $z$ direction, and minima with $E_{wz}$ fields in the negative $z$ direction. As a result the net electric wave force per m$^3$ on the electrons, averaged over time ($F_{we} = -e<n_e E_{wz}>$), is directed in the negative $z$ direction.

(3) With the Debye length of the order of 0,02 mm, and the wavelength of the structure about 30 mm, there is quasineutrality within the waves. The electric wave force on the ions is therefore equal in strength, and opposite in direction to, the force on the electrons, $F_{wi} = q_i<n_i E_{wz}> = e<n_e E_{wz}> = -F_{we}$. The ions experience a force in the positive $z$ direction.

The action of the waves is thus identical to the action of electron-ion collisions in the sense that it represents a momentum exchange between ions and electrons, in the direction of the macroscopic current. The role of electron-ion collisions in magnetic diffusion is to dissipate the current, and allow magnetic field penetration. The waves should therefore have a similar effect. This analogy between the wave structure and ion-electron collisions was carried out in [2,3]. It was demonstrated that the action of the waves could be represented by an effective anomalous transverse resistivity

$$\eta_{EFF} = \eta_c + <n_e E_{wz}>/e(v_{iz} - v_{ez})<n_e>^2) \qquad (1)$$

Where $\eta_c$ is the classical Spitzer resistivity. The quantity $\eta_{EFF}$ was calculated directly from measured $n_e$ and $E_{wz}$ such as in Fig. 6c and 6d. One typical such calculation from measurements in the standard position B [3] gave $\eta_{EFF} = 0,0068\Omega m$ (where the classical resistivity $\eta_c$ contributed only 0,4%), giving a magnetic diffusion coefficient $D_B = \eta_{EFF}/\mu_0 = 5400 m^2 s^{-1}$, and a diffusion time for the magnetic field into the plasma of $\tau_B = \mu_0 L^2/(4\eta_{EFF})$ about 0,5µs. This is what is needed to explain the observed (Fig. 3b) penetration of 70% of $B_z$ during the one microsecond of passage through the magnetic transition.

We want to stress that the conclusions so far regarding the magnetic diffusion due to the observed waves are drawn directly from the measured data, without any assumption of the nature of the waves, or how they are driven. This aspect is discussed in [3] based on measured wavelengths, **k** vector orientation, phase velocity, and the current density (and direction) by which the instability is driven. The saturation to high amplitudes during the transition (Fig 4) gives a measure of the growth rate. Comparisons both with linear analytical theory, and with a PIC computer simulation made for the purpose, led [3] to the conclusion that the closest label is the modified two-stream instability (MTSI), driven by a diamagnetic current that is set up in response to the time-changing magnetic field (*i.e.,* time-changing in the plasma's rest frame).

**4. The hf terms in the generalized Ohm's law**

A useful approach is to use the generalized Ohm's law to obtain the different time scales corresponding to various mechanisms for magnetic field changes [10]. In the form of Spitzer [11] it reads

$$(m_e/e^2 n_e)(d\mathbf{J}/dt) + \eta\mathbf{J} + \mathbf{J}\mathrm{x}\mathbf{B}/en_e = \mathbf{E} + \mathbf{v}\mathrm{x}\mathbf{B} + \mathrm{grad}p/en_e \qquad (2)$$

This equation can be rewritten [10] into an equation for the time rate of change of the magnetic field by using the Maxwell Equations for curl**B** and curl**E**, while neglecting the displacement current. For example, classical ion-electron collisions give a contribution in the resulting equation, which originates in the resistive term $\eta\mathbf{J}$ of Eq. (2). A comparison between this and the convective term (originating in the **v**x**B** term) gives the magnetic Reynolds number, which is convenient to assess whether, in a classically resistive plasma flow, magnetic diffusion dominates over the convection (with the flow) of the magnetic field. Anomalous diffusion of various kinds has often been described in terms of an effective collision frequency, which is equivalent to ascribing the wave (or turbulence) effect directly to the resistive term in Eq. 2. This was the approach in the preceding section. Although it was concluded that the hf waves in our experiment make their action on the electrons via the first term to the right in Eq 2 (the **E** term), the geometry is such (notably $\mathbf{F}_{wi} = -\mathbf{F}_w$, and along the direction of the macroscopic current **J***)* that the effect, when averaged over the wave structure, can conveniently be expressed as an increase in effective resistivity $\eta_{EFF}$. Other authors [9, 10, 12], have associated the anomalous magnetic diffusion with the Hall term $\mathbf{J}\mathrm{x}\mathbf{B}/(en_e)$ in Eq 2. A question of some interest is how these different points of view can be reconciled in a concrete case such as ours: should the anomalous magnetic diffusion best be associated with the resistive term $\eta\mathbf{J}$ of Eq. (2), with the Hall term $\mathbf{J}\mathrm{x}\mathbf{B}/(en_e)$, or with the electric term **E**? We will use a simplified cartoon version (Fig. 7) of the observed wave pattern (Fig. 6) to discuss this issue.

Fig. 7a shows, approximately to scale, a cut in the *xz* plane through the plasma stream in the device at the time when we take the wave hf data, $t = 12 - 14\mu s$. The front of the plasma stream is just about reaching the bottom plate. The plasma stream which had a cylindrical symmetry before the transition, has at $z = 0$ been compressed to a typical width 0,1m in the *x* direction under the action of the diamagnetic current, and expanded to about 0,3m in the *y* direction [1]. At $z = 0$, a diamagnetic current is flowing dominantly in the –*z* direction on the high potential side of the stream (positive *x*) , and in the +*z* direction in the low potential side (negative *x*). It  flows along the sides of the stream and closes at the back and front. Fig. 7b shows an enlargement of the plasma stream around $z = 0$, and Figures 7c and 7d finally show schematically the wave features at the position where they are measured, around $(x, y, z) = (0,03, 0, 0)$. (Notice that the figures 7c and 7d show only the wave features. The larger scale diamagnetic current is not drawn, and neither is the larger scale self-polarization field $\mathbf{E}_p = -\mathbf{v}\mathrm{x}\mathbf{B}$. Fig. 7c and 7d are intended to highlight separately the physics of the waves. For the complete picture, the large-scale fields and currents must be included).

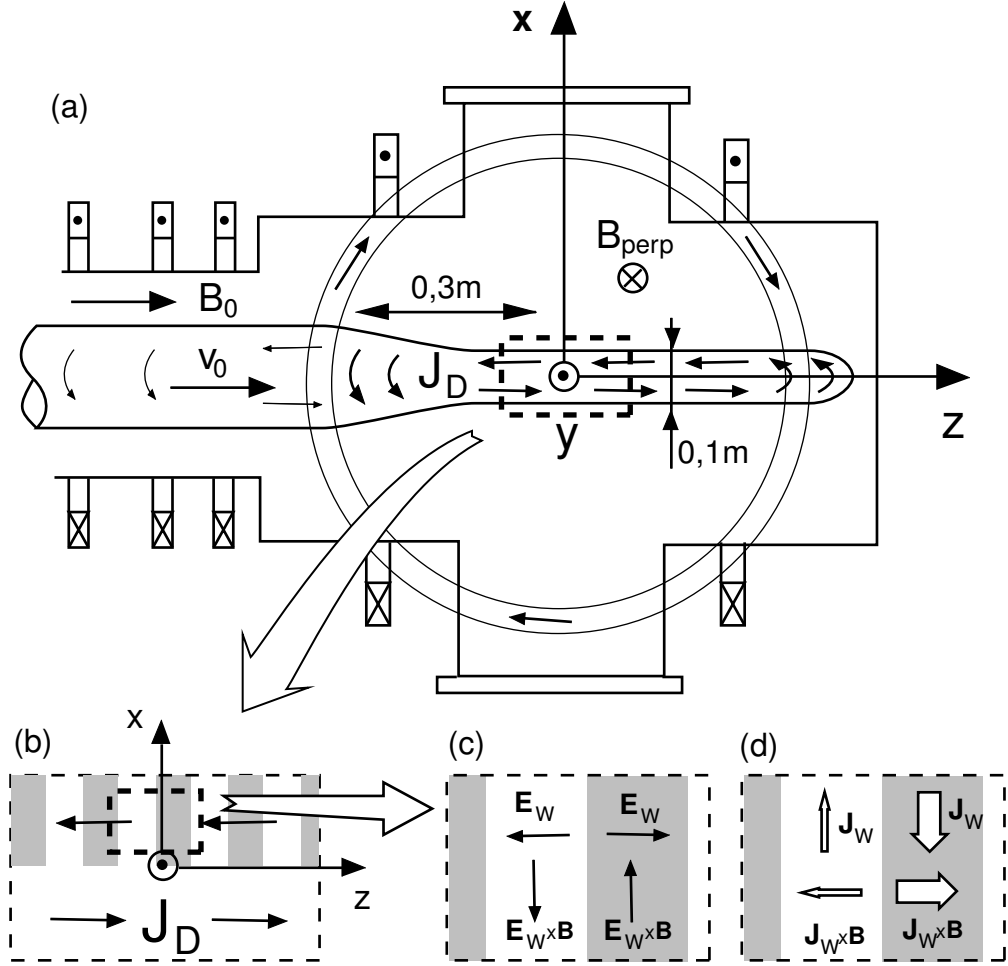

Fig. 7a. The plasma stream at 12 – 14 μs. 7b : a close-up of the centre of the plasma stream. 7c and 7d : the details in the wave pattern.

The generalized Ohm's law in the form of Eq. (2) is valid on the electron time scale. Let us call it the 'high frequency' generalized Ohm's law. It can be applied also to the high frequency features in the waves, and there is no room in it for mechanisms like anomalous resistivity or anomalous collision frequencies. Such concepts need to be introduced if instead a 'low frequency' generalized Ohm's law is used, where the variables in the equation are the averages over the hf features. High frequency features can generally give contributions to most terms in such a low frequency equation. There can be contributions from oscillations in ion velocity **v**, magnetic field **B**, electric field **E**, etc. In our case it is simpler because we can make estimates of the relative amplitudes based on measurements. We find that the dominating hf contributions are in the Hall term (from oscillating currents in the $x$ direction) and in the **E** term (from the oscillating $E_z$ component described in the previous section). Eq. (2) can be rewritten with these terms separated into a quasi-dc (QDC) part and a wave (W) part :

$$\ldots + \mathbf{J}_{QDC} \times \mathbf{B}/(en_e) + \mathbf{J}_W \times \mathbf{B}/(en_e) = \mathbf{E}_{QDC} + \mathbf{E}_W + \ldots \quad (3)$$

The wave field $\mathbf{E}_W$ is mainly along the $z$ direction as described in the previous section. The $\mathbf{J}_W \times \mathbf{B}/(en_e)$ term is obtained from measurements with 3-axis magnetic pickup coils which show hf oscillations with the same spectra as for the electric field and density oscillations in Fig. 5, and with the dominating component corresponding to currents flowing back and forth along the wave fronts, in the *x* direction. No independent measure of their phase velocity has yet been made. However, assuming that it is the same as the phase velocity of the density and electric field oscillations, the spatial structures and the current density can be obtained from the time series of the magnetic oscillations. The result is consistent with a wave-associated current structure as shown in Fig. 7d, where the $\mathbf{J}_W \times \mathbf{B}/(en_e)$ term in Eq. (3) is easily identified. The $\mathbf{J}_W$ current is here due to the $\mathbf{E}_W$ field in Fig. 7c, which gives the electrons a Hall drift in the *x* direction $\mathbf{E}_W \times \mathbf{B}/B^2$. (This drift gives a current $\mathbf{J}_W$ because the electrons, but not the ions, are magnetized on the wave scale since $r_{gi} : k_w^{-1} : r_{ge} = 20 : 1 : 0,1$).

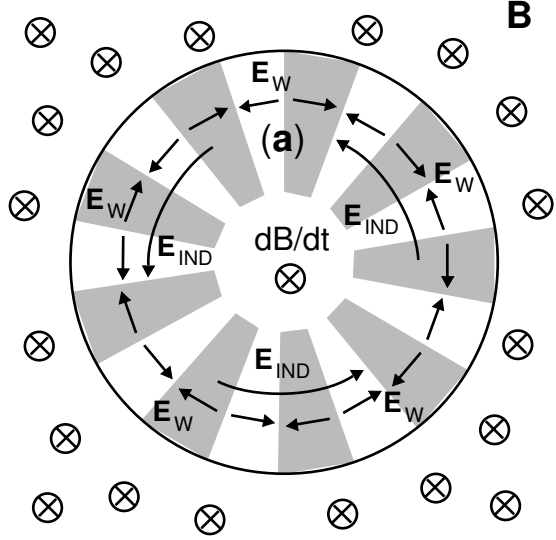

Fig. 8. A cartoon of a simple case of hf-assisted magnetic field diffusion: a stationary cylindrical plasma in an external magnetic field that grows in time. As discussed in the text, this structure enables also completely magnetized and collisionfree electrons to be unfrozen from the magnetic field as it diffuses inwards. If we make a thought experiment where the figure above is stretched out in the horizontal direction, the point (a) in this figure would correspond to the position where we make measurements in the experiment, $(x, y, z) = (0,03, 0, 0)$. The experimental wave field $E_W$ and the density at this point are shown in Fig. 6c, 6d, and in the cartoon of 7c. Since the figure above shows a stationary plasma in a time-changing magnetic field, there is no exact counterpart in the experiment to the induced field $E_{IND}$. However, at the point in the experiment corresponding to point (a) above, the $E_z$ component of the quasi-dc electric field plays the same role. This is discussed in section 3.4 of the companion paper [13].

An example in simpler geometry illustrates more clearly the role of structures such as in Fig. 7c and 7d for magnetic field diffusion. Consider a cylindrical symmetric case of magnetic field diffusion, Fig. 8. The azimuthal **E** field is the sum of the $\mathbf{E}_W$ field from the wave structure and the induced electric field $\mathbf{E}_{IND}$ due to $dB/dt$. Let us take the case where the electrons are neither drifting inwards nor outwards. Their zero velocity in the radial direction can then be seen as the sum of two equally fast, opposing drifts : (1) an inwards drift $\mathbf{E}_{IND} \times \mathbf{B}/B^2$, which corresponds to the electrons being frozen onto the magnetic field in its motion towards the centre, and (2) an (average) outwards drift in the wave pattern. The drift in the wave pattern is actually alternatingly in and out, depending on the sign of the wave field at the azimuthal coordinate in question. The net wave-produced drift of the whole electron population however, is non-zero, given by the average $<\mathbf{E}_w \times \mathbf{B}\, n_e> /(<n_e> B^2)$. With the density and the electric field correlated as in Fig. 8, the wave gives a drift contribution outwards, counteracting the inwards 'drift' with the magnetic field lines.

We are now in a position to return to the question of which is the best way to describe the anomalous magnetic diffusion in our experiment. In the high frequency version of the generalized Ohm's law (Eq 2), the waves give contributions to both the Hall term and the electric term. If an average is taken over the spatial wave structure, there is a balance between a net force on the electrons from the high frequency $\mathbf{E}_W$ term, and a net 'high-frequency-produced' $<\mathbf{J}_W \times \mathbf{B}> = <\mathbf{J}_W> \times \mathbf{B}$ force. The macroscopic volume current $<\mathbf{J}_W>$ due to the waves comes from the ($\mathbf{E}_w \times \mathbf{B}$) Hall drift of the electrons in the hf $\mathbf{E}_w$ field. This *hf Hall drift* gives the electron population an average drift velocity $<\mathbf{J}_W>/en_e$ across $\mathbf{B}$ and can be directly interpreted as in terms of anomalous magnetic diffusion, as in Fig 8. In this view, the Hall term is most directly responsible for the magnetic diffusion. If a low frequency version of the generalized Ohm's law is desired (*i e*, one where $\mathbf{E}$ and $\mathbf{B}$ are the quasi-dc electric fields), the wave action can in our case (where these forces are along the macroscopic diamagnetic current $\mathbf{J}$) be formally attributed to the resistive term in the equation. Since we have reliable measurements of the wave $\mathbf{E}_w$ term (from data such as in Fig. 6) we use those to calculate the anomalous resistivity $\eta_{EFF}$. This approach gives anomalous magnetic diffusion in terms of an anomalous resistivity. In this view, an anomalous ion-electron collision frequency can be regarded as responsible for the anomalous magnetic diffusion. In our experiment, it is of the order of $10\omega_{lh}$ [2,3]. We regard these two points of view (magnetic diffusion due to the Hall term, or due to anomalous resistivity) as equivalent for the case of our experiment. Furthermore, it is only a question of measurements that we use the $\mathbf{E}$ term for the calculation of $\eta_{EFF}$. If we instead had had more reliable measurements of the wave current $\mathbf{J}_W$ in the Hall term, we could have based the expression for the anomalous resistivity on those measurement. A goal for future experiment is to check that these two ways of estimates give consistent results.

In [4], and in a companion paper in this issue [13], the results from the present experiment are generalized to other parameter regimes with the goal to determine from basic parameters what is likely to happen when a plasmoids meets a magnetic barrier: penetration through magnetic expulsion, penetration through self-polarization, or rejection.

**Acknowledgements**

This work was supported by the Swedish Research Council.